# Characterizing a fiber-based frequency comb with electro-optic modulator.

W. Zhang[1], M. Lours[1], M. Fischer[2], R. Holzwarth[2], G. Santarelli[1] and Y. Le Coq[1]

[1]*LNE-SYRTE, Observatoire de Paris, CNRS, UPMC, 61 avenue de l'Observatoire, Paris,*

*France*

[2]*MenloSystems GmbH, Am Klopferspitz 19a, D-82152 Martinsried, , Germany*

**Abstract**

We report on the characterization of a commercial-core fiber-based frequency comb equipped with an intracavity electro-optic modulator (EOM). We investigate the relationship between the noise of the pump diode and the laser relative intensity noise (RIN) and demonstrate the use of a low noise current supply to substantially reduce the laser RIN. By measuring several critical transfer functions, we evaluate the potential of the EOM for comb repetition rate stabilization. We also evaluate the coupling to other relevant parameters of the comb. From these measurements we infer the capabilities of the femtosecond laser comb to generate very low phase noise microwave signals when phase locked to a high spectral purity ultra-stable laser.

*1.Introduction.*

Femtosecond lasers have revolutionized the field of time and frequency metrology by providing a phase coherent link between a large span of optical and microwave frequencies [1-7]. These so called optical frequency combs are widely used for optical frequency measurements with uncertainty close to the limitations of the caesium fountain clocks [8-11], as well as frequency comparisons between different optical frequency standards [12,13]. By transferring the spectral purity of ultra stable cw-lasers [14-18] to the microwave domain with minute excess added noise [19-21], they also provide ultra-low phase noise microwave signals [22-24]. Such signals may be of wide technological application in fields such as radar, telecommunications, deep space navigation systems, timing distribution and synchronization [25]. They have proven to be suitable as interrogation signal for atomic fountain clocks at the quantum projection noise limit [26,27].

Among the different mode-locked laser technologies available, the Titanium Sapphire (Ti:S) laser has traditionally been the workhorse of most optical metrology laboratories. Now Erbium-doped fiber-based systems are currently evolving into very serious contenders, especially when high reliability and very long term operation (several days or weeks of operator-free continuous measurements) are crucial aspects. The main drawbacks of Erbium-doped fiber-based systems, when compared to Ti:S lasers, have been the higher noise and lower control bandwidth available. Laboratory systems with an intra-cavity Electro-Optic Modulator (EOM) for improved control bandwidth have been demonstrated [28-30]. Highly reliable systems, including an EOM actuator and exhibiting the high repetition rates suitable for optical frequency metrology, are now available commercially with a built-in self-referencing unit [31]. We present in this paper several characterizations of such a commercial-core laser and demonstrate how we have used this information to understand and minimize the residual noise of the system when phase locked to an ultra-stable cavity stabilized cw laser, with a strong emphasis on low phase noise microwave signal generation. We start by presenting the various parameters and actuators of the commercial-core optical frequency comb. We proceed with a description of the methods used to measure actuators response (transfer functions) and the relevant noise properties of our comb. We conclude by applying these studies to the low phase noise microwave generation.

*2. Optical frequency comb*

A self referenced optical frequency comb is characterized, in the frequency domain by a set of phase coherent optical frequencies $\nu_N = N \times f_{rep} + f_0$, where $N$ is the index of the mode, $f_{rep}$ the repetition rate (typically hundreds of MHz) and $f_0$ the carrier-envelope offset frequency [1]. A free-running optical frequency comb exhibits fluctuations of both $f_{rep}$ and $f_0$, as well as of its average amplitude $A$. The state of the laser at a given time is therefore characterized by these three distinct parameters. Note that, in principle, beyond these three "global parameters" one could imagine that the phase and amplitude relation between the different spectral components of the comb could also be changing over time. Due to the non-linear effects responsible for the mode-locking mechanism, such variations have been shown to be minute and, at best, have a very slow evolution [32]. We will therefore not take such effect into account in this paper. A full characterization of the comb's noise properties therefore requires the measurement of three power spectral densities (PSD), of either phase or frequency noise for $f_0$ and $f_{rep}$, and of amplitude noise for $A$. Note that the effect of parameter $A$ is usually neglected in optical frequency metrology. In the context of low phase noise microwave signal generation, where amplitude-to-phase coupling in the photodetection process is a strong limitation [33,34], characterizing the amplitude fluctuations of the laser is however of utmost interest.

These three parameters ($f_0$, $f_{rep}$ and $A$), can be controlled via multiple actuators. In this paper, we characterize a femtosecond comb system which possesses a piezo electric actuator (PZT, whose purpose is to change the laser cavity length), allows fine tuning of the laser pumping via adjustment of the current which drives the pump laser diodes, and has an EOM in the cavity. The latter element operates mainly as a very fast voltage-controlled group delay element in the femtosecond laser's cavity. It therefore changes the repetition rate, although with unavoidable coupling to the other two parameters ($f_0$ and $A$). Ideally, the PZT modifies only $f_{rep}$. However it is well known that a tiny misalignment of the laser cavity may influence both $f_0$ and $A$. Changing the pump power of the laser comb (*via* the current driving the pump diode lasers) obviously impacts $A$, as well as both $f_0$ and $f_{rep}$ via complex mode-locked laser dynamics [35-37]. A full characterization of the response of the comb to the various actuators therefore requires the measurement of the nine transfer function matrix elements (three actuators for three comb's parameters) as shown by the following equation:

$$\begin{pmatrix} A(j\omega) \\ f_{rep}(j\omega) \\ f_0(j\omega) \end{pmatrix} = \begin{bmatrix} H_{I_{pump},A}(j\omega) & H_{V_{PZT},A}(j\omega) & H_{V_{EOM},A}(j\omega) \\ H_{I_{pump},f_{rep}}(j\omega) & H_{V_{PZT},f_{rep}}(j\omega) & H_{V_{EOM},f_{rep}}(j\omega) \\ H_{I_{pump},f_0}(j\omega) & H_{V_{PZT},f_0}(j\omega) & H_{V_{EOM},f_0}(j\omega) \end{bmatrix} \begin{pmatrix} I_{pump}(j\omega) \\ V_{PZT}(j\omega) \\ V_{EOM}(j\omega) \end{pmatrix} \quad (1)$$

where $I_{pump}$ is the pump diode current $V_{PZT}$ is the voltage driving the PZT and $V_{EOM}$ is the voltage driving the EOM crystal. We have measured the nine transfer functions of this matrix but we will focus on the subset which impact optical frequency metrology and optical-to-microwave division process.

We proceed with a description of the methods we developed to measure these transfer functions, as well as several noise properties of the femtosecond laser.

*2. Transfer functions and noise measurements techniques.*

Our commercial-core fiber-based frequency comb has a repetition rate near $f_{rep}$~250MHz, which can be coarsely adjusted to within +/- 1MHz using a motorized translation stage on one of the cavity mirrors. The available optical output power is about 130 mW. The laser has a built-in *f-2f* interferometer unit which generates the offset frequency signal $f_0$. A motorized double wedge [37] allows coarse frequency adjustment of this quantity, which we maintain near 70 MHz. To measure transfer functions, we use a two channel vector signal analyzer (VSA - Agilent 89410A) which can operate up to 10 MHz Fourier frequency and has a with built-in programmable source voltage output. The applied modulation is directed simultaneously to the Channel 1 input of the VSA and the actuators of the laser which we want to characterize (PZT, pump diode current or EOM) via a suitable actuator driver. Note that these drivers need to be independently characterized, to insure they don't impose bandwidth limitation to the measurement. The quantity to characterize ($f_{rep}$, $f_0$ or $A$) is transformed into a voltage signal and fed to Channel 2 of the VSA. By programming the VSA source output to generate a chirped sine wave, we measure the transfer function from Ch1 to Ch2.

This allows obtaining the transfer function from the actuator to the comb parameter. The set-up schematic is depicted in figure 1.

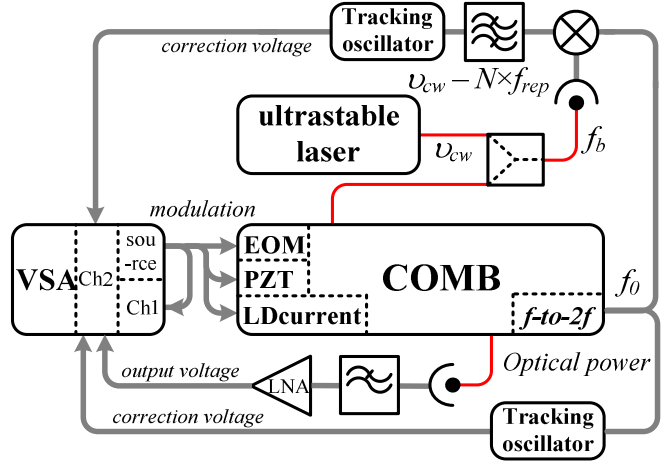

Fig 1: Setup schematic used to measure transfer functions of amplitude, $f_0$ and $f_{rep}$ vs laser pump current, intra cavity EOM voltage and PZT voltage, as well as power spectral densities of the same quantities. *f*-to-2*f*: built-in self-referencing unit, LNA: low noise amplifier, VSA: vector signal analyzer.

The key part of such an experiment is to design ways to transform $f_{rep}$, $f_0$ and $A$ into a voltage signal with high sensitivity and minimal cross talk from the other two parameters of the comb. Measuring $A$ is quite straightforward. We inject a few milliwatts of light from the femtosecond laser output (30 nm bandwidth around 1.55 µm) onto a fiber-pigtailed InGaAs photodiode. The output of the photodiode is low pass filtered to remove the harmonics of the repetition rate and, after low noise amplification, the output voltage is fed to the VSA. To measure $f_0$ fluctuations, we process this signal with a home-made fast tracking oscillator filter. This device is composed of a voltage-control oscillator (VCO), which is phase locked to the input signal with more than 3 MHz of bandwidth. At Fourier frequencies lower than this bandwidth, the correction signal of the tracking oscillator (*i.e.* the voltage which controls the VCO when the phase lock loop is operating) is proportional to the frequency fluctuations of $f_0$ (the free running VCO noise is negligible compared to that of $f_0$). To measure $f_{rep}$ fluctuations, it is convenient to increase the sensitivity by measuring $N \times f_{rep}$, with $N$ a large integer. By beating the comb output with an ultra stable laser (linewidth<1Hz) [17] of optical frequency $\nu_{cw}$ (near the 1.55 µm central wavelength of the comb), we have access to $f_b = N \times f_{rep} + f_0 - \nu_{cw}$. This radio-frequency (RF) signal mixed with $f_0$ leads to two sidebands, highly sensitive to $f_{rep}$ fluctuations (because of the large multiplicative factor $N$). We select, with a bandpass filter, the sideband which is independent from $f_0$ (since $f_b - f_0 = N \times f_{rep} - \nu_{cw}$). Locking a tracking oscillator to this signal therefore leads to a voltage correction signal proportional to the fluctuations of $f_{rep}$.

We use these techniques to transform $A$, $f_0$ and $f_{rep}$ into voltage signals to measure the nine transfer functions and the noise power spectral densities of the optical frequency comb.

*3. Results and applications*

Once the comb is characterized, by the 3x3 transfer functions matrix, we can use this information to improve the comb's properties. Our work puts a strong focus on the context of low phase noise microwave signal generation by photo-detection of the repetition rate's harmonics when the comb is phase locked to an ultra stable cw laser [23,24,27,38]. However, the use of the transfer functions matrix goes well beyond what as appears in the following.
A first example of the use of the transfer function/noise characterization is illustrated in fig.2.

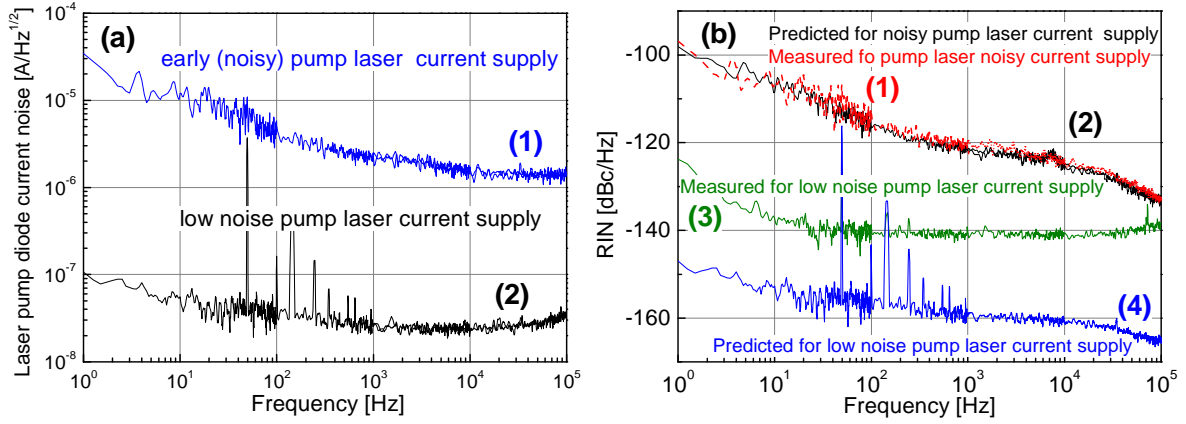

Fig 2: Plot (a): measured a spectral density of current noise for early noisy (1) (blue line) and low noise (2) (black line) current supply. Plot (b): RIN of the femtosecond laser output with two different current supplies for the pump lasers diodes. Curve (1) red (measured) and (2) black (predicted) are for the early (noisy) current supply. Curve (3) green (measured) and curve (4) blue (predicted) are for the low noise current supply.

The first comb was equipped with a relatively noisy pump diode lasers current supply (Fig. 2 plot (a), blue curve (1)). By plugging the current noise of this current supply to the transfer function we measured from the pump diodes' current to the amplitude of the laser, we obtain a predicted relative intensity noise (RIN) of the comb which matches very well the actual measurement (Fig 2 plot (b), red curve (1) & black curve (2)). We can conclude from this that the rather important noise of the early prototype current supply originates the excessive RIN of the laser comb. Indeed, by using a low noise current supply as a replacement (Fig. 2 plot (a), black curve (2)), we obtained a very substantial improvement of the RIN. Furthermore, by comparing, for the low noise current supply, the predicted RIN (Fig. 2 plot (b), green curve (3)) to the new RIN measurement (Fig. 2 plot (b), blue curve (4)), we can conclude that the new current supply is not a limiting factor anymore.

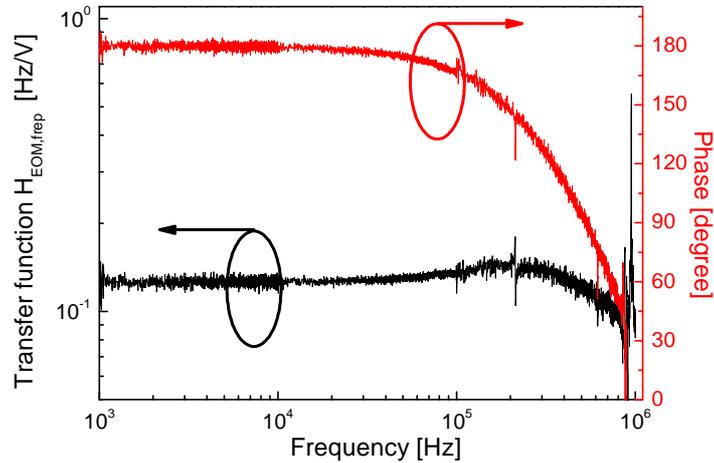

Fig 3: Transfer function $H_{V_{EOM}, f_{rep}}$ from EOM voltage to repetition rate $f_{rep}$. Black is amplitude (left axis), red plot is phase (right axis). The ultimate bandwidth available with such actuator is limited by the strong narrow resonances near 900 kHz Fourier frequency.

By studying the RIN of the pump laser diode itself (when driven by the low-noise current supply), we observed a RIN higher than expected from the current supply's characteristic alone. This excess noise explains the comb's RIN for Fourier frequencies lower than 30-50 kHz. Improving the performance may require to hand-select lower RIN pump laser diodes and/or actively servo their output power. For Fourier frequencies higher than 30-50 kHz, the RIN of the pump diode lasers does not explain RIN of the comb. Understanding this excess noise at high Fourier frequencies would necessitate further theoretical and experimental studies.

A second example of transfer function application is to optimize the feedback loop when phase locking the comb to an external reference (an ultra stable 1.542 µm laser in our case).We give, in fig 3 and 6, the three transfer functions from the EOM control voltage to, respectively $f_{rep}$, $f_0$ and $A$.

From Fig 3, we deduce that the maximum available servo bandwidth when using the EOM as an actuator to control $f_{rep}$ is limited by sharp resonances observed around 900 kHz. As a matter of facts, they produce rapid large phase shifts so large that they cannot be easily compensated by control circuit design. We use proportional and multiple integration control which acts on both the EOM and the PZT actuators (both act mainly on $f_{rep}$) to phase lock $f_{rep}$ to a cw ultra stable laser near 1.55 µm. Similarly to the technique we previously described for measuring $f_{rep}$ fluctuations, the beatnote between the cw laser and one nearby tooth of the comb is mixed with $f_0$ to provide a signal of frequency $N \times f_{rep} - \nu_{cw}$, independent of $f_0$. By mixing it with a reference from a fixed synthesizer, we produce an error signal which is sent to a proportional-integral (PI) controller. This PI controller directly steers the EOM. The EOM control voltage is further integrated and fed to the PZT. In this way $f_{rep}$ is phase-locked to the optical frequency $\nu_{cw}$, with the EOM acting for Fourier frequencies larger than 6-10 kHz, while the PZT controls the laser for low Fourier frequencies. Fig. 4 (a) shows the in-loop error spectral density and Fig. 4 (b) the correction voltage spectral density applied to the EOM when the phase-lock loop is running.

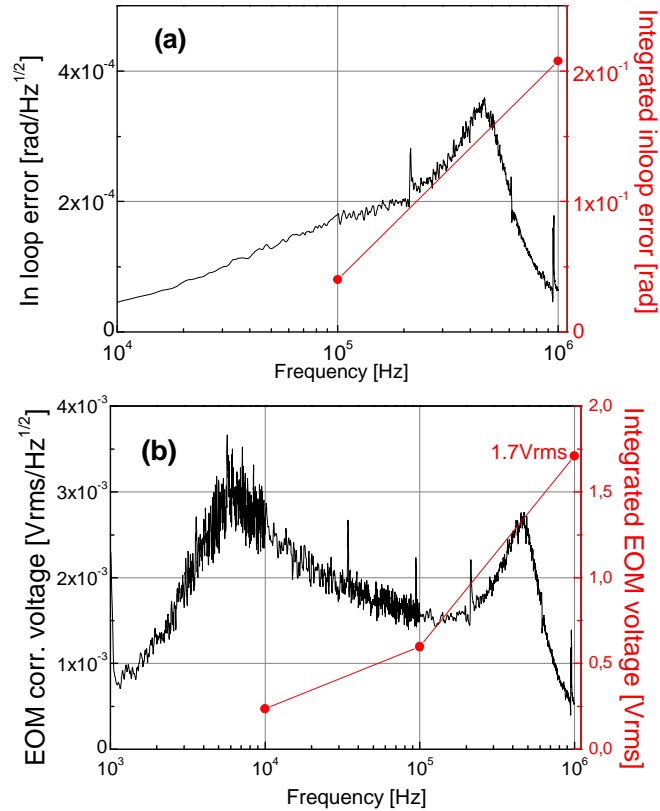

Fig 4: Plot (a): Inloop residual phase error signal when phase locking the comb to a continuous ultra stable laser near 1.55 µm. In red (right axis) the integrated phase error signal. Plot (b): rms correction voltage applied to the EOM when the phase lock loop is running. In red (right axis), the integrated rms voltage applied to the EOM. The decrease of the correction voltage below 6 kHz is due to the use of the PZT controlling the laser for low Fourier frequencies (in lieu of the EOM).

The inloop error exhibits a resonance around 900 kHz and a servo bump near 400 kHz,

consistent with the transfer function presented in fig. 3. By integrating the in-loop error up to 1 MHz, the total phase error is about 0.21 rad rms, resulting in 96.7% energy in the carrier. As the total phase error is much lower than 1 rad rms, the comb is in the so called "ultra-stable regime", where each tooth of the comb has nearly the same spectral purity as the ultra-stable cw laser used as a reference for the lock (once $f_0$ is removed).

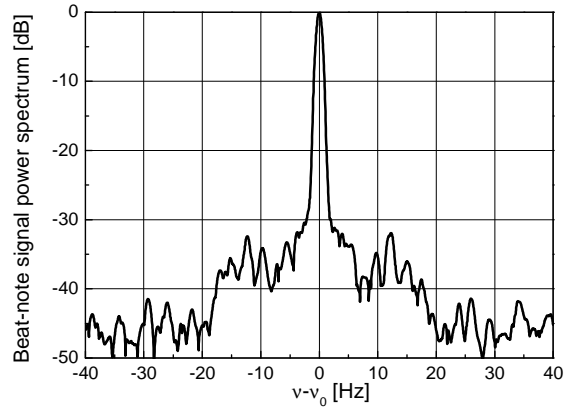

Fig 5: Radio frequency spectrum of the beat- note signal between the stabilized comb and an ultra-stable laser near 1062.5 nm (Span 80 Hz, RBW 1Hz, 10 averages)

This was confirmed independently by beating the comb output near 1062.5 nm (obtained from the same highly non-linear fiber used in the built-in *f-2f* unit) with a second ultra stable laser (see Fig. 5)[39,40]. The integrated correction voltage is 1.7 Vrms (~6 Vpp maximum, verified with an oscilloscope). When comparing to the 25 V output range of our fast control electronics, this leaves room to optimize the EOM dimensions, using shorter crystals with lower dispersion and, potentially, higher resonances which may allow larger control bandwidths. A third example of the use of the transfer function is directly related to the coupling transfer functions that are presented in Fig. 6. This figure represents the unavoidable cross talks to, respectively, $f_0$ and *A*, when we act on the EOM to control $f_{rep}$ (and, ideally, $f_{rep}$ only). Understanding the dynamics of these two cross-talk responses (enhanced responses at high Fourier frequencies in particular) is not straight-forward. We believe them to be caused by minute misalignments of the EOM crystal, which couples, via complex laser dynamics, to the polarization and amplitude of the pulse in the femtosecond laser's cavity. Verifying this hypothesis would require a thorough theoretical and experimental analysis which is well beyond the scope if this paper. The measured transfer functions of Fig. 6 provide some useful information about the limit of the EOM (in its current implementation) as an actuator to phase lock $f_{rep}$.

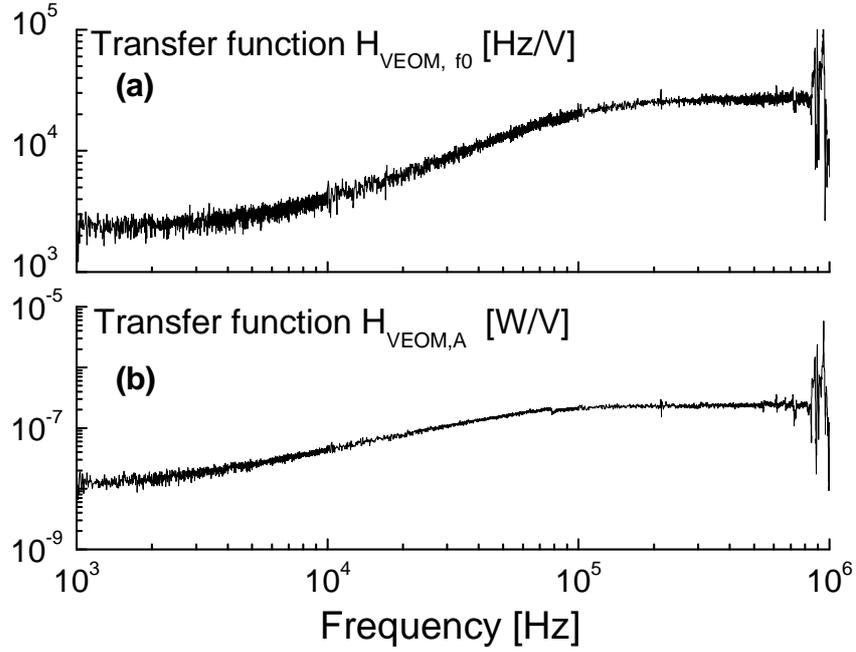

Fig 6: Plot (a) transfer function $H_{V_{EOM}, f_0}$ (EOM voltage to $f_0$). Plot (b) $H_{V_{EOM}, A_0}$ (EOM voltage to comb's amplitude)

By combining the correction voltage spectral density applied to the EOM when the servo loop operating (Fig. 4(b)) to the transfer functions of Fig 6, we can deduce the EOM control-induced fluctuations of $f_0$ and $A$. The fluctuations of $f_0$ are not important in the context of our scheme, where $f_0$ is uncontrolled and simply removed from $f_b$ by frequency mixing. Other $f_0$-removal techniques, including the one presented in [41], are likewise inherently immune to such cross talk. Note that in experiments which independently phase lock $f_{rep}$ and $f_0$ as in Refs [8,42,43]) may however need to account for those effects. On the other hand, the cross-talk induced fluctuations of $A$ are of pre-eminent importance in the context noise microwave signal extraction by photo-detection of the femtosecond pulses. Indeed, amplitude-to-phase conversion which occurs in the photo-detection process limits the achievable performance of low phase noise microwave generation experiments [33, 34]. From the predicted cross-talk-induced fluctuations of $A$, and a given amplitude-to-phase conversion factor related to the photodetection process, one can deduce the achievable level of residual microwave phase noise. For instance, in our case, for a microwave signal at 12 GHz generated from a 250 MHz repetition rate laser, we assume a $d\phi/(dP/P) \sim 0.03$ rad amplitude-to-phase conversion factor of the photodetection process. Here, $\phi$ is the phase of the microwave compared to that of the pulse train and P is the average optical power on the photodetector. Note that substantially higher conversion factors are normally without the fine adjustment of the experimental parameters [33,34].

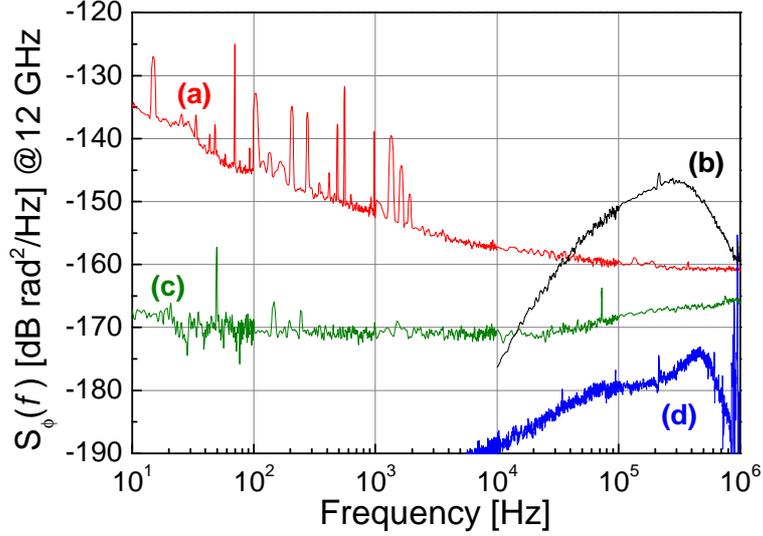

Fig. Residual phase noise for optical-to-microwave frequency division operation of the frequency comb at 12 GHz. Curve (a) (red line): detection noise floor limit (shot noise at high Fourier frequencies) from ref [44], plot (b) (black line): phase noise induced by EOM inloop error (i.e. due to limited bandwidth of the EOM actuator). Plot (c) (green line) : RIN-induced phase noise. Plot (d) (blue line): EOM voltage to amplitude induced phase noise. To obtain curves (c) and (d), we assumed typical 0.03 rad per relative amplitude change of the amplitude-to-phase conversion factor in the photodetection process.

Fig. 7 presents the predicted residual phase noise limits obtained from RIN-induced phase noise (in green) and EOM cross-talk-induced phase noise (in blue) value of the amplitude-phase conversion factor. When comparing these predictions to the best noise floor limit we measured for such a system with a repetition rate multiplication technique detailed in [44], it appears that the cross-talk is not a limiting factor at the present level of noise floor limit. Note however that if the amplitude-phase conversion factor were equal to 1 rad (typical worst case scenario observed from [38]), it would constitute a limitation for Fourier frequencies in the 10 kHz-1 MHz range substantially higher than the measured noise floor limit. Furthermore, the RIN-induced residual phase noise is also lower than the noise floor limit. Similarly, an amplitude-phase conversion factor higher than 0.1 rad would induce a substantial limitation to the performances of the system. On the other hand, it also appears from Fig. 7 that the main limitation in the 100 kHz-1 MHz Fourier frequencies range is due to the residual inloop error of the phase lock loop (black line). This limitation is ultimately linked to the still limited actuation bandwidth achievable with the EOM. Further improvement of the EOM actuator would therefore be necessary to achieve residual phase noise limits in the -160 dB(rad$^2$/Hz). Note that recently a 1.2 MHz bandwidth was achieved with a waveguide EOM [45] and we could therefore expect a similar result for the future prototypes.

## 4. Conclusions

We have presented in detail the techniques used to measure the various transfer functions which characterize an optical frequency comb equipped with various actuators. We have shown some examples of how to use such characterization to deduce the performance limits of the comb and improve its performance. When phase locking the repetition rate of the comb to an ultra stable cw laser optical frequency, we reached the "ultra-stable regime", where the linewidth of each comb tooth is limited by that of the cw laser reference (and not the phase-lock loop performances). In the context of low phase noise microwave generation by photodetecting the pulse train, we have made a prediction of the achievable minimum phase noise from different possible noise sources. These prediction make us confident that we can achieve residual phase noise in the -160 dB(rad$^2$/Hz) level, close to the shot noise limit, with a slightly improved EOM actuator. In addition we have shown how to use the dynamic characterization of a frequency comb (*via* its actuators transfer functions) to understand and improve an optical frequency comb's performance.

*5. Acknowledgments*

We want to thanks J. Pinto for the valuable work for the electronic sub systems, and John Mc Ferran for careful reading of the manuscript.